\documentclass[fleqn,twoside]{article}
\usepackage{espcrc2}
\usepackage{graphicx}

\newcommand{\lsim}{\raisebox{-4pt}{$
\,\stackrel{\textstyle <}{\sim}\,$}}

\newcommand{\gev}{\mbox{~GeV}}
\newcommand{\mev}{\mbox{~MeV}}

\newcommand{\half}{{\textstyle\frac{1}{2}}}

\newcommand{\tvec}[1]{\mbox{\boldmath{$#1$}}}
\newcommand{\svec}[1]{\mbox{\boldmath{$\scriptstyle #1$}}}

\newcommand{\slim}{\mskip 1.5mu}
\newcommand{\abst}{|\slim t\slim|}

\title{Generalized parton distributions from form factors%
\thanks{To appear in: Procs.\ of the Workshop on Light-Cone QCD and
Nonperturbative Hadron Physics 2005 (LC 2005), Cairns, Australia, 2005}
}

\author{M. Diehl\,\address{Deutsches Elektronen-Synchroton DESY, 22603
    Hamburg, Germany}}

\begin{document}

\begin{abstract}
The electromagnetic nucleon form factors provide constraints on
generalized quark distributions.  Key results of the study presented
here are a strong dependence of the average impact parameter of quarks
on their longitudinal momentum fraction, a striking difference in the
$t$ dependence of $u$ and $d$ quark contributions to elastic form
factors, and an estimate of the orbital angular momentum carried by
valence quarks in the nucleon.
\vspace{1pc}
\end{abstract}

\maketitle

%%%%%%%%%%%%%%%%%%%%%%%%%%%%%%%%%%%%%%%%%%%%%%%%%%%%%%%%%%%%%

\section{INTRODUCTION}
\label{sec:intro}

\begin{figure}[htb]
\begin{center}
\includegraphics[width=0.3\textwidth]{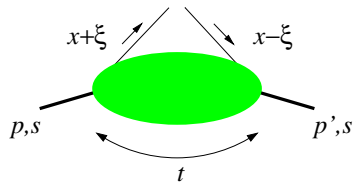}
\vspace{-3.5em}
\end{center}
\caption{\label{fig:gpd} Relevant variables in a GPD.}
\end{figure}

Much of what we know about hadron structure comes from measurements of
parton densities, which quantify the distribution of longitudinal
momentum and helicity of partons in a fast-moving hadron.  Generalized
parton distributions (GPDs) complement this essentially
one-dimensional picture with information in the plane perpendicular to
the direction of movement.  These distributions parameterize matrix
elements of non-local quark or gluon operators.  We focus here on the
unpolarized quark sector, where the distribution $H^q(x,\xi,t)$ is
diagonal in proton helicity and $E^q(x,\xi,t)$ describes proton
helicity flip.  The variables $x$ and $\xi$ parameterize longitudinal
quark momentum fractions relative to the average proton momentum
$\half (p+p')$ as shown in Fig.~\ref{fig:gpd}, whereas the invariant
$t=(p-p')^2$ depends on both longitudinal and transverse components of
the momentum transferred to the proton.  In the forward limit $p=p'$
one recovers the usual quark and antiquark densities as $q(x) =
H^q(x,0,0)$ and $\bar{q}(x) = -H^q(-x,0,0)$ with $x>0$.  For the
precise definitions of GPDs and further information we refer to the
recent reviews~\cite{Diehl:2003ny}.

According to factorization theorems, GPDs appear in the scattering
amplitudes of suitable hard exclusive processes such as deeply virtual
Compton scattering, $\gamma^* p\to \gamma\slim p$, and exclusive meson
production, e.g.\ $\gamma^* p\to \rho\slim p$.  In these processes,
the longitudinal momentum transfer $\xi$ is fixed by the kinematics,
whereas $x$ is a loop variable.  To disentangle the $x$ and $\xi$
dependence of GPDs from measured process amplitudes remains an
outstanding task, but generically the dominating values of $x$ in the
loop integrals will be of order~$\xi$.

%%%%%%%%%%%%%%%%%%%%%%%%%%%%%%%%%%%%%%%%%%%%%%%%%%%%%%%%%%%%%

\section{IMPACT PARAMETER DENSITIES}
\label{sec:impact}

To represent GPDs in transverse position space, we form wave packets
\begin{equation}
  \label{impact-state}
|p^+, \tvec{b}\rangle = \int\frac{d^2\tvec{p}}{(2\pi)^2}\,
  e^{-i\svec{b} \svec{p}}\, |p^+, \tvec{p} \rangle
\end{equation}
from momentum eigenstates $|p^+, \tvec{p} \rangle$, where we write
$v^{\pm} = (v^0 \pm v^3) /\sqrt{2}$ for the light-cone components and
$\tvec{v} = (v^1, v^2)$ for the transverse part of a four-vector $v$.
The state $|p^+, \tvec{b}\rangle$ is localized at $\tvec{b}$ in the
transverse plane (often called impact parameter plane).  Formally it
is an eigenstate of a suitably defined transverse position operator
\cite{Soper:1972xc}.  It is thus possible to localize a relativistic
state exactly in \emph{two} dimensions, whereas localization in all
\emph{three} dimensions involves ambiguities at the level of the
Compton wavelength.  For a parton interpretation it is natural to
consider states $|p^+, \tvec{b}\rangle$ with large $p^+$, which
describe a fast-moving proton.  Further analysis reveals that
$\tvec{b}$ is the ``center of momentum'' of the partons in the proton,
given as $\tvec{b} = \sum_i p_i^+ \tvec{b}_i^{\phantom{+}} /\sum_i
p_i^+$ in terms of their plus-momenta and transverse positions.  The
center of momentum is related by Noether's theorem to transverse
boosts, in analogy to the relation between the center of mass and
Galilean transformations in nonrelativistic mechanics.

Taking matrix elements between impact parameter states
(\ref{impact-state}) of the quark or gluon operators defining general
parton distributions in momentum space, one obtains Fourier transforms
of these distributions.  For vanishing skewness parameter $\xi$ one
finds that
\begin{equation}
q(x,\tvec{b}) = \int \frac{d^2\tvec{\Delta}}{(2\pi)^2}\,
  e^{-i \svec{b} \svec{\Delta}}\,
  H^q(x,0,-\tvec{\Delta}^2)
\end{equation}
is the density of quarks with longitudinal momentum fraction $x$ and
transverse distance $\tvec{b}$ from the center of momentum of the
proton \cite{Burkardt:2000za}.  For nonzero $\xi$ one no longer has a
probability interpretation because the two quark momentum fractions in
Fig.~\ref{fig:gpd} are not the same, but $\tvec{b}$ still describes
the distribution of the struck quark in the transverse plane
\cite{Diehl:2002he}.  According to the discussion at the end of the
introduction, the combined $\xi$ and $t$ dependence of hard exclusive
scattering processes thus yields information about the impact
parameter distribution of partons with longitudinal momentum fraction
of order $\xi$.  Note that the connection between GPDs and spatial
distributions in the transverse plane discussed here differs from the
well-known representation of form factors in terms of spatial
distributions in \emph{three} dimensions, which has been extended to
GPDs in~\cite{Belitsky:2003nz}.

\begin{figure}
\begin{center}
\includegraphics[width=0.25\textwidth]{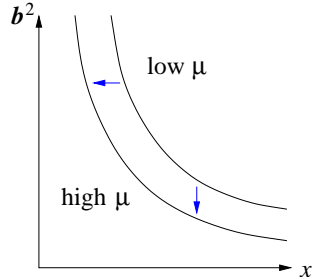}
\vspace{-3em}
\end{center}
\caption{\label{fig:evol} Typical pattern of scale evolution in $x$
and $\tvec{b}^2$.}
\end{figure}

Just as the ordinary parton densities, the distributions
$q(x,\tvec{b})$ depend on the scale $\mu$ at which the partons are
resolved.  The scale evolution is local in $\tvec{b}$ and described by
the usual DGLAP equations.  For the valence quark distributions
$q_v(x,\tvec{b}) = q(x,\tvec{b}) - \bar{q}(x,\tvec{b})$ we have
\begin{equation}
\mu^2 \frac{d}{d\mu^2}\, q_v(x,\tvec{b}) 
= \int_x^1 \frac{dz}{z}\,
  \Big[ P\Big(\frac{x}{z}\Big) \Big]_+\, q_v(z,\tvec{b}) \,,
\end{equation}
where $P(z)$ denotes the quark splitting function.  As a consequence
the $\tvec{b}$ dependence of the distributions at given $x$ changes
with $\mu$.  In particular, the average squared impact parameter
$\langle \tvec{b}^2 \rangle_x$ for valence quark distributions evolves
as \cite{Diehl:2004cx}
\begin{eqnarray}
  \label{dglap-impact}
\lefteqn{
\mu^2 \frac{d}{d\mu^2}\, \langle \tvec{b}^2 \rangle_x 
}
\\
&=& \hspace{-0.8em} {}-\frac{1}{q_v(x)} \int_x^1 \frac{d z}{z}\,
  P\Big(\frac{x}{z}\Big)\, q_v(z) 
  \Big[ \langle \tvec{b}^2 \rangle_x 
      - \langle \tvec{b}^2 \rangle_z \Big] \,,
\nonumber
\end{eqnarray}
where we defined
\begin{eqnarray}
  \label{av-b}
\langle \tvec{b}^2 \rangle_x
&=& \frac{\int d^2\tvec{b}\; \tvec{b}^2\,
  q_v(x,\tvec{b})}{\int d^2 \tvec{b}\; q_v(x,\tvec{b})} 
\nonumber \\
&=& 4\, \frac{\partial}{\partial t} \log H_v^q(x,t) \Big|_{t=0}
\end{eqnarray}
with $H_v^q(x,t) = H^q(x,0,t) + H^q(-x,0,t)$ being the momentum space
counterpart of $q_v(x,\tvec{b})$.  The evolution equations for singlet
distributions mix quarks and gluons as usual.  We will argue below
that $\langle \tvec{b}^2 \rangle_x$ is a decreasing function of $x$.
Since $P(z) > 0$, the average impact parameter at given $x$ then
decreases with $\mu$ according to (\ref{dglap-impact}).  This is
readily understood: at fixed $\tvec{b}$ evolution to higher scale
$\mu$ decreases the longitudinal momentum of quarks because they
radiate gluons.  As sketched in Fig.~\ref{fig:evol}, this implies a
smaller typical $\tvec{b}$ at given $x$ as $\mu$ increases.

\begin{figure}[tb]
\begin{center}
\includegraphics[width=0.38\textwidth]{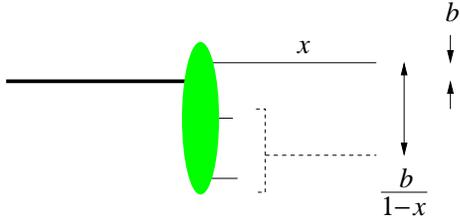}
\vspace{-3em}
\end{center}
\caption{\label{fig:geometry} Three-quark configuration with one fast
  quark in a proton.  The thick line denotes the center of momentum of
  the proton and the dashed line the center of momentum of the two
  spectator quarks.}
\end{figure}

At large $x$, the struck quark takes most of the proton momentum, so
that its impact parameter tends to coincide with the center of
momentum of the entire proton.  In the limit $x\to 1$ one thus expects
a narrow distribution in $\tvec{b}$, or equivalently a flat $t$
dependence of GPDs in momentum space.  An estimate for the overall
transverse size of the proton in that limit is provided by the
transverse distance $\tvec{b} /(1-x)$ between the struck quark and the
center of momentum of the \emph{spectator} partons, as shown in
Fig.~\ref{fig:geometry}.  It is plausible to assume that this distance
remains finite due to confinement \cite{Burkardt:2004bv}, so that the
average squared impact parameter of partons vanishes like $\langle
\tvec{b}^2 \rangle_x \sim (1-x)^2$ for $x\to 1$.

%%%%%%%%%%%%%%%%%%%%%%%%%%%%%%%%%%%%%%%%%%%%%%%%%%%%%%%%%%%%%

\section{THE DIRAC FORM FACTORS}
\label{sec:dirac}

Information on the interplay between $x$ and $t$ in the valence
distributions $H_v^q(x,t)$ can be obtained from the Dirac form factors
of proton and neutron via the sum rules
\begin{eqnarray}
  \label{p-sum}
F_1^{p}(t) &=& \textstyle \int_0^1 dx\, \Big[
   \frac{2}{3} H_v^{u}(x,t) 
 - \frac{1}{3} H_v^{d}(x,t) \Big] ,
\\
  \label{n-sum}
F_1^{n}(t) &=& \textstyle \int_0^1 dx\, \Big[
   \frac{2}{3} H_v^{d}(x,t) 
 - \frac{1}{3} H_v^{u}(x,t) \Big] ,
\end{eqnarray}
where we have neglected the contribution from strange quarks.  Flavor
labels in $H_v^q$ refer to quarks in the proton.  Notice that the
dependence of GPDs on the resolution scale $\mu$ cancels in these
integrals, because the electromagnetic form factors belong to a
conserved current.  In this way, form factors measured at low $t$ can
constrain the distributions of partons resolved at much higher
resolution scales $\mu^2$.

The information from elastic form factors is very complementary to
what can be learned from hard exclusive scattering processes.
Experimental coverage in $t$ and the precision of measurements and
their quantitative interpretation is typically greater for form
factors than for more complex exclusive processes.  Electromagnetic
form factors are sensitive to the difference of quark and antiquark
distributions and thus insensitive to sea quarks and gluons, which can
be accessed in processes like deeply virtual Compton scattering or
vector meson production.  Finally, parton momentum fractions only
appear under an integral in elastic form factors, whereas a combined
measurement of the $\xi$ and $t$ dependence in exclusive processes
gives a more direct correlation of longitudinal and transverse
variables as discussed in Sect.~\ref{sec:impact}.

%%%%%%%%%%%%%%%%%%%%%%%%%%%%%%%%

\subsection{An ansatz for the distribution $H_v^q$}
\label{sec:fit}

In the following we present results of the analysis of electromagnetic
form factors performed in Ref.~\cite{Diehl:2004cx}, to which we refer
for details.  A study along similar lines can be found in
\cite{Guidal:2004nd}.  Our ansatz for $H_v^q$ is of the form
\begin{equation}
  \label{H-ansatz}
H_v^q(x,t) = q_v(x) \exp[\, t f_q(x) \,] ,
\end{equation}
where for $q_v(x)$ we have taken the CTEQ6M parameterization
\cite{Pumplin:2002vw}.  The results of our analysis are stable within
the CTEQ error estimates on the parton densities.  All distributions
in the following refer to the scale $\mu=2 \gev$.  The interplay
between $x$ and $t$ dependence in $H_v^q$ is controlled by the profile
function $f_q(x)$, which according to (\ref{av-b}) is readily
identified as $\frac{1}{4} \langle \tvec{b}^2 \rangle_x$.

\begin{figure*}
\begin{center}
\includegraphics[width=.4\textwidth,
  bb=120 235 500 590]{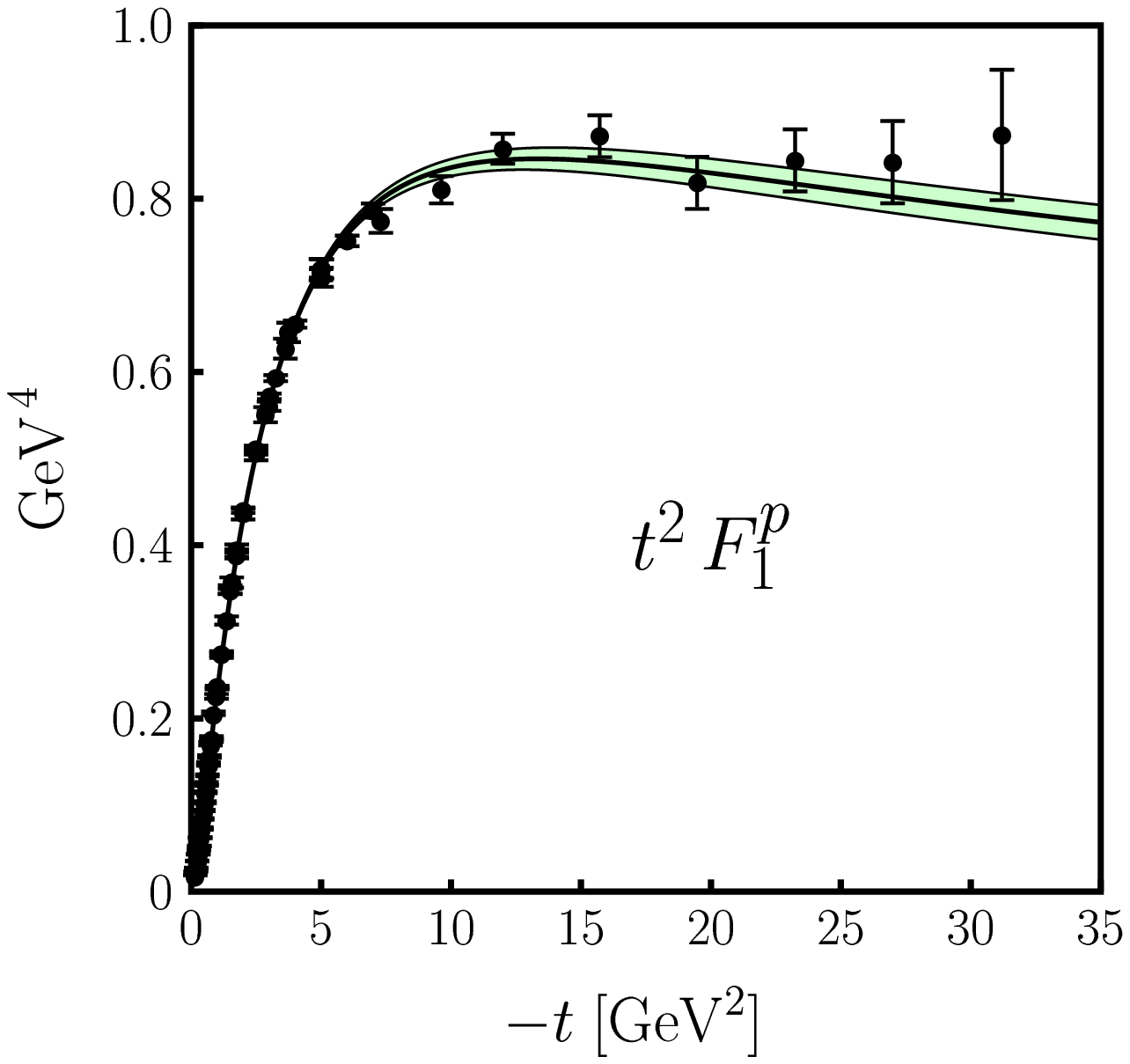}
\hspace{2em}
\includegraphics[width=.4\textwidth,
  bb=110 337 490 690]{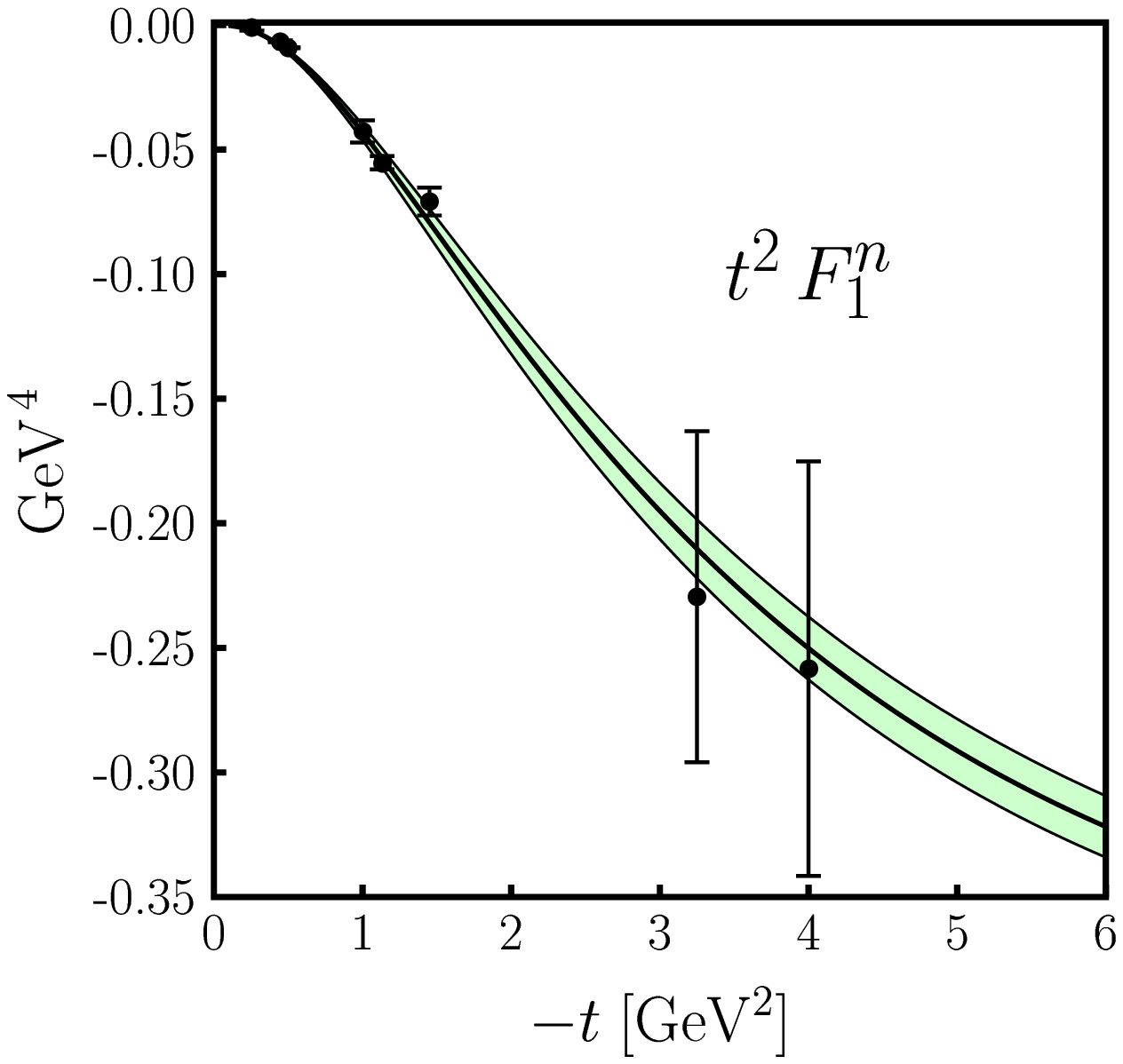}
\vspace{-3em}
\end{center}
\caption{\label{fig:F1fit} Results of fitting the ansatz given by
  (\protect\ref{H-ansatz}) and (\protect\ref{profile-ansatz}) to the
  proton and neutron Dirac form factors using the sum rules
  (\protect\ref{p-sum}) and (\protect\ref{n-sum}).  Shaded bands
  reflect the 1--$\sigma$ uncertainties on the fitted parameters.  The
  data for $F_1^p$ are described within 5\% except for the point at
  the highest $\abst$.}
\end{figure*}

For small $x$, Regge phenomenology of soft hadronic interactions
suggests an ansatz \cite{Goeke:2001tz}
\begin{equation}
  \label{regge-ansatz}
H_v^q(x,t) \;\sim\;  x^{-(\alpha+\alpha' t)}
  \;=\;  x^{-\alpha} \, e^{\,t \alpha' \log(1/x)}
\end{equation}
for the $x$ dependence, with $\alpha\approx 0.4 \mbox{~to~} 0.5$ and
$\alpha'\approx 0.9 \gev^{-2}$ corresponding to the leading meson
exchange trajectories.  This is known to work rather well for the
usual valence quark densities, i.e.\ in the forward limit $t=0$.  We
note that in the singlet sector the situation is more complicated: the
powers $\alpha$ parameterizing sea quark and gluon densities at scales
of a few GeV are significantly larger than the respective values for
meson and pomeron exchange in soft hadronic reactions.  Furthermore,
the value of $\alpha'$ measured in exclusive $J/\Psi$ production,
which involves the generalized gluon distribution, is smaller than the
corresponding value for pomeron exchange in soft scattering processes
\cite{Chekanov:2004mw}.  It remains an outstanding task to constrain
the shrinkage parameter $\alpha'$ in generalized sea quark
distributions and to understand its interplay with $\alpha'$ for
gluons through evolution in $\mu$.

For the limit $x\to 1$ we impose $f_q(x) \sim (1-x)^2$, corresponding
to a finite transverse size of the proton as discussed in
Sect.~\ref{sec:impact}.  We investigated several forms of $f_q$ that
interpolate between the limiting behavior for $x\to 0$ and $x\to 1$
just discussed, and found good results with
\begin{eqnarray}
  \label{profile-ansatz}
f_q(x) &=& \alpha' (1-x)^3 \log(1/x) 
\nonumber \\
 && {}+ B_q (1-x)^3 + A_q x (1-x)^2 .
\end{eqnarray}
The high power of $(1-x)$ multiplying the term with $\log(1/x)$
ensures that the parameter $\alpha'$ controls the behavior of the
distribution at small but not at moderate or large $x$ -- we expect
that the physics in these $x$ regions is very different and not
naturally described by the same parameters.

%%%%%%%%%%%%%%%%%%%%%%%%%%%%%%%%

\subsection{Results and lessons of the fit}

\begin{figure}
\begin{center}
\includegraphics[width=0.45\textwidth]{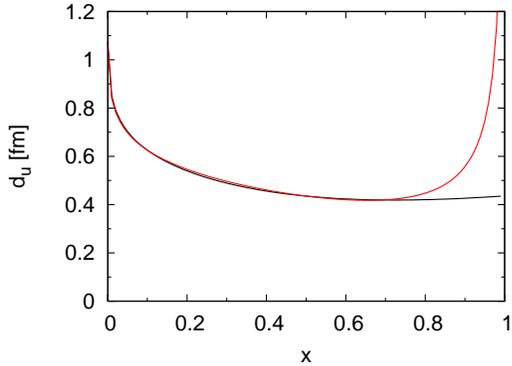}
\vspace{-3em}
\end{center}
\caption{\label{fig:profile} Average distance between struck quark and
  spectators in the valence distribution for $u$ quarks, given by $d_u
  = (1-x)^{-1}\, [\langle b^2 \rangle_x ]^{1/2}$ according to
  Fig.~\protect\ref{fig:geometry}.  The lower curve is for our fit
  (\protect\ref{profile-ansatz}) and the upper one for the alternative
  fit (\protect\ref{alt-ansatz}).}
\end{figure}

With the ansatz just described we obtain a good fit to the data for
the Dirac form factors, as seen in Fig.~\ref{fig:F1fit}.  To reduce
the number of free parameters we set $\alpha' =0.9 \gev^{-2}$; leaving
it free we find $\alpha' =0.97\pm 0.04 \gev^{-2}$ well in the region
suggested by Regge phenomenology.  We also imposed $B_u=B_d\slim$;
relaxing this constraint improves the fit only slightly.  The main
result of our fit is a strong $x$ dependence of the average impact
parameter of valence quarks over the entire $x$ range.  This is
illustrated for $u$ quarks in Fig.~\ref{fig:profile}, where we plot
the average distance $d_u$ between struck quark and spectators.

\begin{figure}[t]
\begin{center}
\includegraphics[width=0.4\textwidth,%
  bb=60 320 410 565]{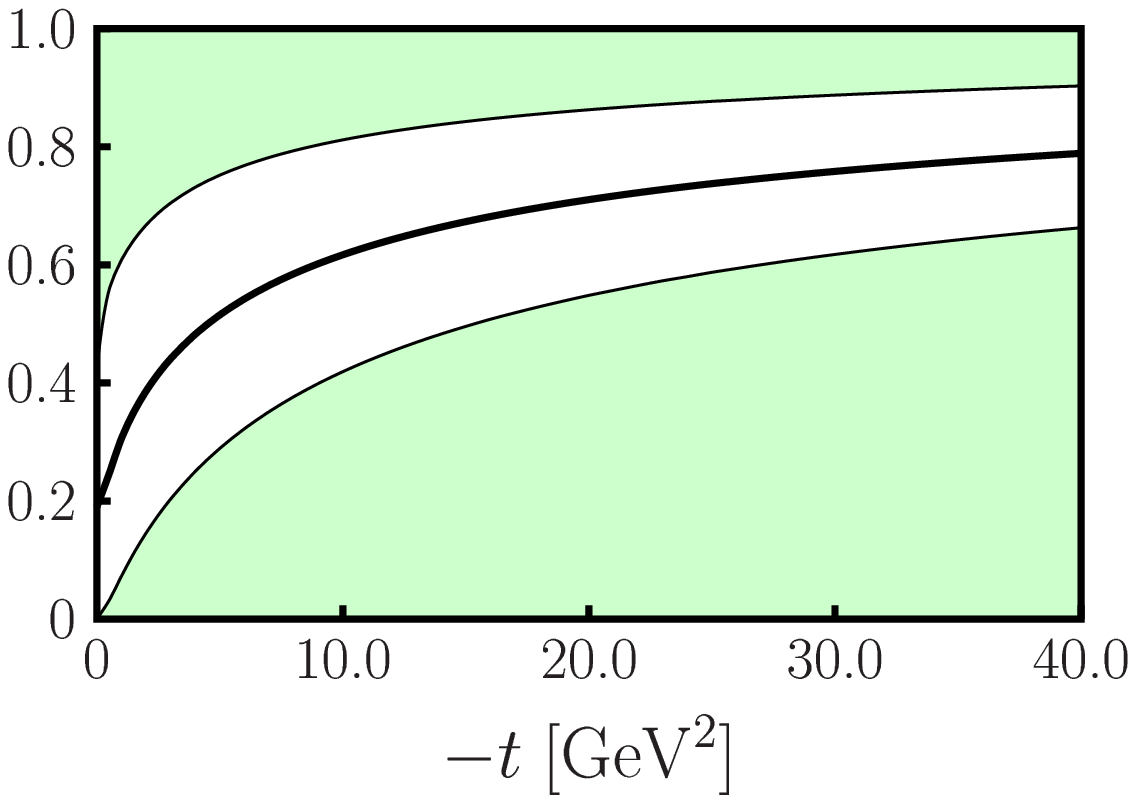}
\vspace{-3.4em}
\end{center}
\caption{\label{fig:minmax} Region of $x$ (white region) which
accounts for $90\%$ of $F_1^p(t)$ in the integral
(\protect\ref{p-sum}) for our fit of $H_v^q(x,t)$.  The upper and
lower shaded $x$-regions each account for $5\%$ of $F_1^p(t)$.  The
thick line shows the average $\langle x\rangle_t$ as defined in
(\protect\ref{xavg}).}
%\end{figure}
%
\vspace{0.7em}
%
%\begin{figure}
\begin{center}
\includegraphics[width=0.45\textwidth]{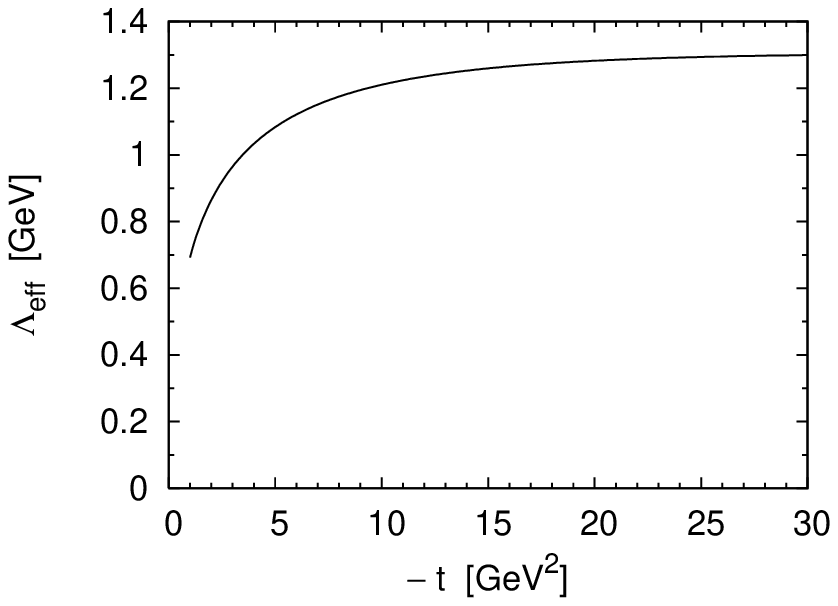}
\vspace{-3em}
\end{center}
\caption{\label{fig:leff} The scale parameter $\Lambda_{\rm eff}$ for
  the Feynman mechanism, as explained in the text.}
\end{figure}

We have also fitted the form factors to the alternative ansatz
\begin{eqnarray}
  \label{alt-ansatz}
f_q(x) &=& \alpha' (1-x)^2 \log(1/x) 
\nonumber \\
 && {}+ B_q (1-x)^2 + A_q x (1-x)
\end{eqnarray}
for the profile function.  This leads to a similarly good description
of the data as with the form (\ref{profile-ansatz}).  The reason can
be seen in Fig.~\ref{fig:profile}.  Up to $x \sim 0.8$ the fitted
results for $d_u(x)$ are barely distinguishable; only for larger $x$
does the fit with (\ref{alt-ansatz}) produce a rise of $d_u(x)$ to
values that appear unphysically large.  In the $t$ range where there
is data, the form factor $F_1^p$ is however barely sensitive to
$x>0.8$.  This is shown in Fig.~\ref{fig:minmax}, where we also plot
the average value
\begin{equation}
  \label{xavg}
\langle x \rangle_t = \frac{\sum_q e_q \int_0^1 dx\, x
  H_v^q(x,t)}{\sum_q e_q \int_0^1 dx\, H_v^q(x,t)}
\end{equation}
of $x$ in the integral (\ref{p-sum}).  The more scarce data on $F_1^n$
do not constrain the region $x>0.8$ either.  We see from this exercise
that with observables sensitive to $x \lsim 0.8$ one cannot
unambiguously determine a power-law behavior in $(1-x)$ for the limit
$x\to 1$.

The average impact parameter for $d$ quarks is less well constrained
by our fit since $F_1^p$, for which data is abundant, is dominated by
$u$ quarks.  We found however that a good description of the $F_1^n$
data requires a larger impact parameter of $d$ quarks compared with
$u$ quarks at moderate to large $x$.  It will be interesting to see
whether future data on $F_1^n$ confirm this trend.  This would be an
analog to the very different distribution in $x$ of $u$ and $d$
quarks, which may for instance hint at a quark-diquark structure of
nucleon configurations at large $x$ \cite{Feynman:1972}.

%%%%%%%%%%%%%%%%%%%%%%%%%%%%%%%%

\subsection{Large $t\,$: Feynman and Drell-Yan}

If one assumes that elastic form factors are dominated by
configurations where partons not struck by the photon have
virtualities of the order of a strong interaction scale $\Lambda^2$,
then at large $t$ the momentum fraction of the struck quark must
become large -- this is the mechanism originally proposed by Feynman
\cite{Feynman:1972}.  The integrals (\ref{p-sum}) and (\ref{n-sum})
are then dominated by the region where $1-x \sim \Lambda /\sqrt{-t}$.
The large-$t$ asymptotics of our ansatz (\ref{H-ansatz}) and
(\ref{profile-ansatz}) indeed follows this behavior, as is readily
seen from the saddle point approximation of the relevant integrals.
To quantify this we have evaluated $\Lambda_{\rm eff}(t) = \sqrt{-t}\;
\langle 1 - x \rangle_t$ with $\langle 1 - x \rangle_t = 1 - \langle x
\rangle_t$ from (\ref{xavg}).  The result is displayed in
Fig.~\ref{fig:leff} and shows that the expected asymptotic behavior
slowly sets in for $\abst$ around $10 \gev^2$.

To explore the dependence of our conclusions on the assumed form of
$H_v^q(x,t)$, we have performed fits with the exponential $t$
dependence in (\ref{H-ansatz}) replaced by a power law
\begin{equation}
H_v^q(x,t) = q_v(x) \,\Big( 1 - \frac{t f_q(x)}{p} \,\Big)^{-p}
\end{equation}
with $f_q(x)$ of the form (\ref{profile-ansatz}).  We obtain a good
description of the form factors in a wide range of $p$, from $p
\approx 2.5$ up to the limit $p\to \infty$, where we recover the
exponential (\ref{H-ansatz}).  Asymptotically, our modified ansatz
still satisfies $\langle 1 - x \rangle_t \sim \Lambda /\sqrt{-t}$.
For small $p$ this behavior is however not reached at values of $t$
where there is data, and the form factor sum rules are not dominated
by large $x$.  (This underlines the need to check asymptotic
considerations against numerical estimates when describing baryon form
factors.)  The available data for $F_1^p$ does hence \emph{not prove}
that the Feynman mechanism is at work in the high-$t$ region, but it
is \emph{consistent} with this assumption, given the success of our
fits with large $p$.

\begin{figure*}[t]
\begin{center}
\includegraphics[width=0.46\textwidth]{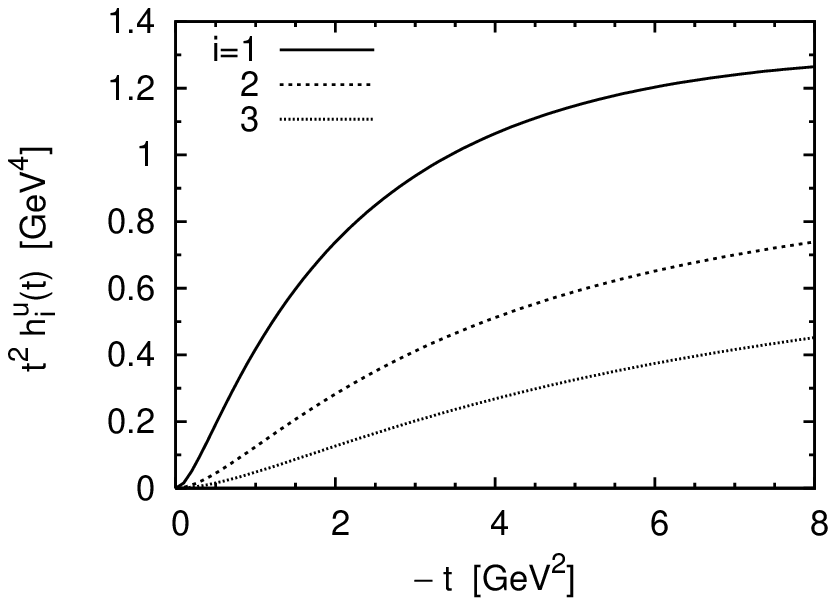} \hfill
\includegraphics[width=0.46\textwidth]{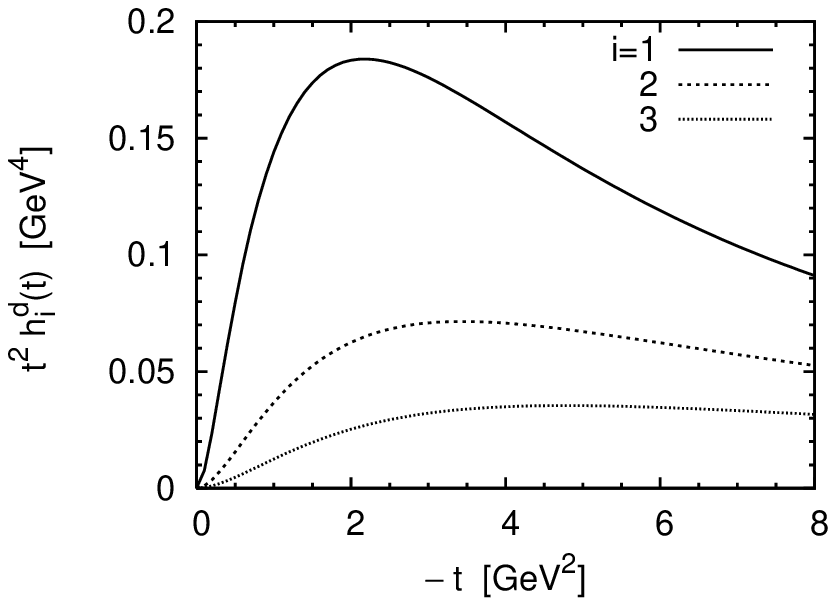} \\
\vspace{1em}
\includegraphics[width=0.46\textwidth]{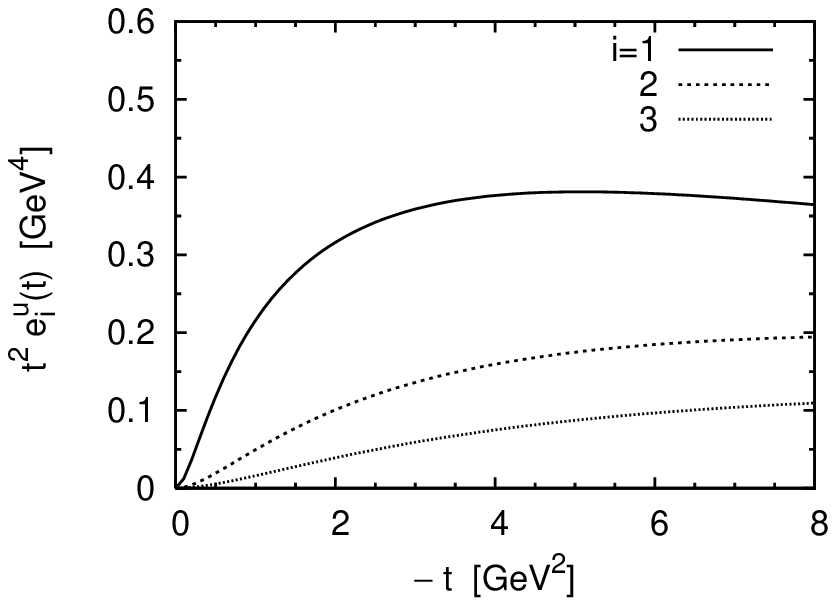} \hfill
\includegraphics[width=0.46\textwidth]{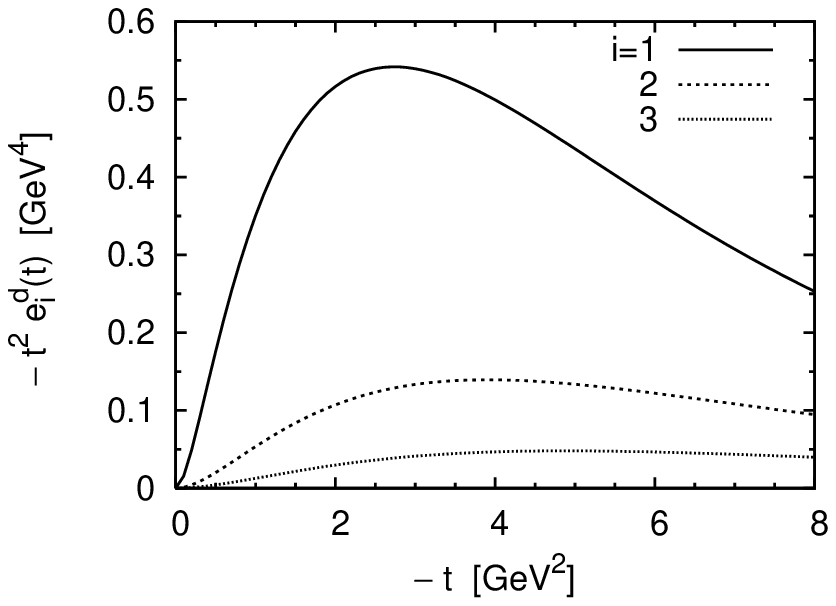} \\
\vspace{-3em}
\end{center}
\caption{\label{fig:moments} Scaled $x$ moments
  (\protect\ref{h-moments}) and (\protect\ref{e-moments}) of
  generalized $u$ and $d$ valence distributions at $\mu=2\gev$,
  obtained from our fits described in Sects.~\protect\ref{sec:fit} and
  \protect\ref{sec:pauli}.}
\end{figure*}

Dominance of the Feynman mechanism implies the Drell-Yan relation
\begin{equation}
  \label{dy}
F_1^q(t) \sim \abst^{- (1+\beta_q)/2}
~~\mbox{for}~~ q_v(x) \sim (1-x)^{\,\beta_q}
\end{equation}
between the form factors $F_1^q(t) = \int dx\, H_v^q(x,t)$ at large
$t$ and parton distributions at large $x$.  We can understand
$\beta_q$ as an \emph{effective} power describing the behavior of
$q_v(x)$ at large $x$ (rather than in the experimentally unexplored
limit $x\to 1$).  With the CTEQ6M distributions at $\mu=2\gev$ we find
$u_v(x) \sim (1-x)^{3.4}$ and $d_v(x) \sim (1-x)^{5.0}$ for $0.5\le x
\le 0.9$.  The Drell-Yan relation then implies a drastically different
$t$ dependence of the contributions from $u$ and $d$ quarks to the
nucleon form factors.  $F_1^u = 2F_1^p + F_1^n$ should approximately
scale like $\abst^{-2}$ at large $t$, whereas $F_1^d = F_1^p + 2F_1^n$
should fall off like $\abst^{-3}$.  This difference in the large-$t$
behavior is already felt at lower values of $t$, as can be seen in the
two upper panels of Fig.~\ref{fig:moments}, where we plot the $x$
moments
\begin{equation}
  \label{h-moments}
h^q_{i}(t) = \textstyle \int_{0}^1 dx \, x^{i-1} \, H_v^q(x,t)
\end{equation}
with $i=1,2,3$ for the GPDs obtained in our fit.  The lowest moments
$h^u_1$ and $h^d_1$ can be extracted from experimental data on the
electromagnetic proton and neutron form factors, and the higher
moments are accessible to calculation in lattice QCD
\cite{Gockeler:2003jf,Hagler:2003jd,Schierholz:2005}.  Both types of
studies are challenging for $\abst$ above $3 \gev^2$, where our fit
predicts the most striking differences between $u$ and $d$ quarks, but
will hopefully be feasible in the future.

%%%%%%%%%%%%%%%%%%%%%%%%%%%%%%%%%%%%%%%%%%%%%%%%%%%%%%%%%%%%%

\section{THE PAULI FORM FACTORS}
\label{sec:pauli}

\begin{figure}
\begin{center}
\includegraphics[width=0.25\textwidth]{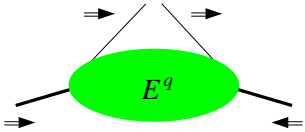}
\vspace{-3em}
\end{center}
\caption{\label{fig:gpdE} $E^q$ describes transitions
   where the proton helicity is flipped but the quark helicity
   conserved.  The helicity mismatch is compensated by one unit of
   orbital angular momentum, according to angular momentum
   conservation.}
\end{figure}

The proton helicity flip distributions $E^q$ admit a density
interpretation at $\xi=0$, similar to the distributions $H^q$
discussed so far.  To see this one changes basis from proton helicity
states $|\!\!\uparrow \rangle$, $|\!\!\downarrow \rangle$ to states
$|X\pm \rangle = (\, |\!\!\uparrow \rangle \pm |\!\!\downarrow \rangle
\,) /\sqrt{2}$ polarized along the positive or negative $x$ axis.  In
impact parameter space one then obtains the density
\begin{equation}
  \label{trans-density}
q_X(x,\mathbf{b}) = q(x,\tvec{b}) 
  - \frac{b^y}{m} \frac{\partial}{\partial \tvec{b}{}^2}\,
    e^q(x,\tvec{b})
\end{equation}
of unpolarized quarks in a proton polarized in the positive $x$
direction, where $q(x,\tvec{b})$ and
\begin{equation}
e^q(x,\tvec{b}) = \int \frac{d^2\tvec{\Delta}}{(2\pi)^2}\,
  e^{-i \svec{b} \svec{\Delta}}\,
  E^q(x,0,-\tvec{\Delta}^2)
\end{equation}
depend on $\tvec{b}$ only through $\tvec{b}^2$ due to rotation
invariance.  The impact parameter distribution of quarks in a
transversely polarized proton is thus shifted in the direction
perpendicular to the polarization.  The interpretation of
(\ref{trans-density}) as a density implies positivity bounds for
$e^q(x,\tvec{b})$, which in momentum space can be written as
\cite{Burkardt:2003ck}
\begin{eqnarray}
  \label{e-bound}
\lefteqn{ \Big[ E^q(x,0,t=0) \Big]^2 }
\\
&\leq &  m^2 
\Big[ q(x) + \Delta q(x) \Big]\, \Big[ q(x) - \Delta q(x) \Big]\,
\nonumber \\
 && {}\times 4\, \frac{\partial}{\partial t} 
    \log \Big[H^q(x,0,t)\pm \tilde H^q(x,0,t)\Big]_{t=0} \;.
\nonumber
\end{eqnarray}
Depending on the sign, the term in the third line is the average
squared impact parameter of quarks with positive or negative helicity,
which according to our previous discussion tends to zero for large
$x$.  In addition, the densities $d + \Delta d$ and $u - \Delta u$ of
right-handed $d$ and left-handed $u$ quarks are phenomenologically
known to decrease strongly with $x$, so that the right-hand side of
(\ref{e-bound}) restricts $|E^q(x,0,0)|$ quite severely for larger
values of $x$.  The distribution $E^q$ involves one unit of orbital
angular momentum since in the associated matrix elements the proton
helicity is flipped but the quark helicity conserved, see
Fig.~\ref{fig:gpdE}.  The bound (\ref{e-bound}) thus limits the amount
of orbital angular momentum that can be carried by quarks with large
$x$.

\begin{figure*}[t]
\begin{center}
\includegraphics[width=0.4\textwidth,%
  bb=50 50 398 291]{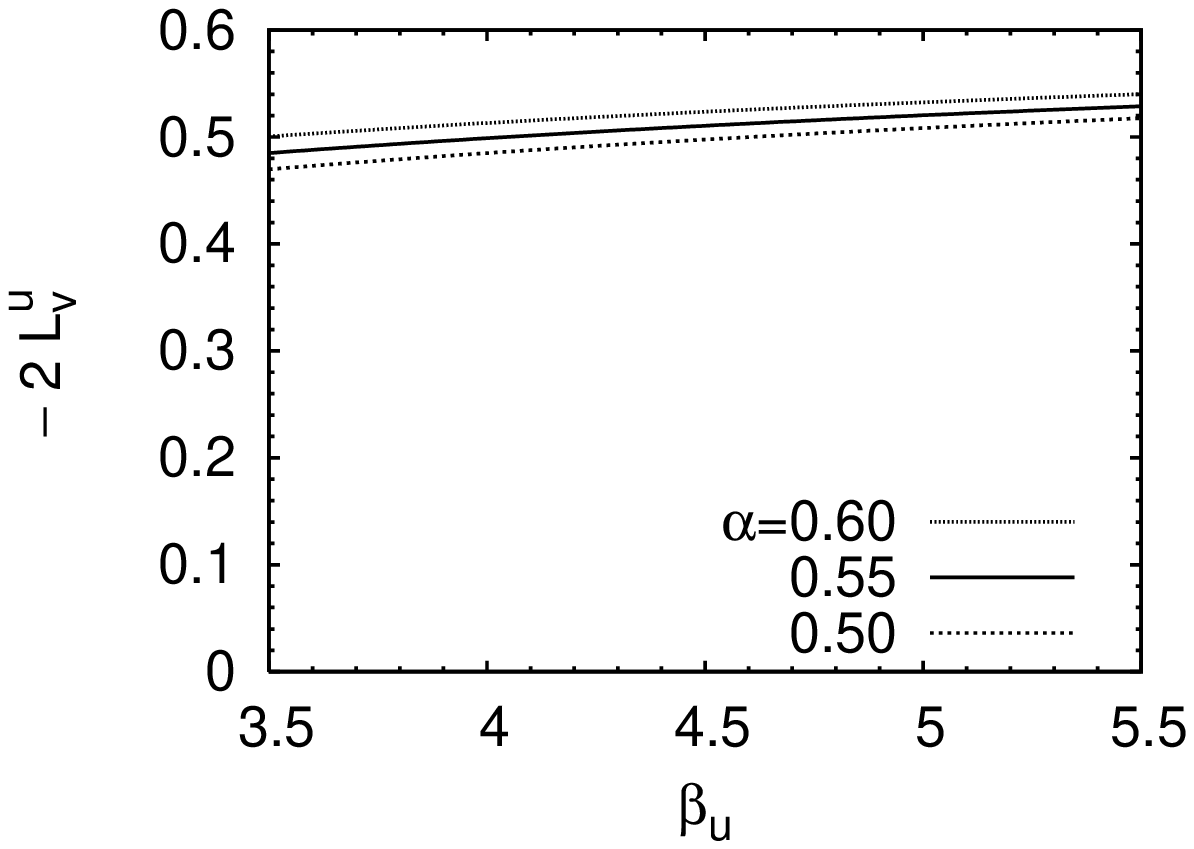}
\hspace{2.5em}
\includegraphics[width=0.4\textwidth,%
  bb=50 50 398 291]{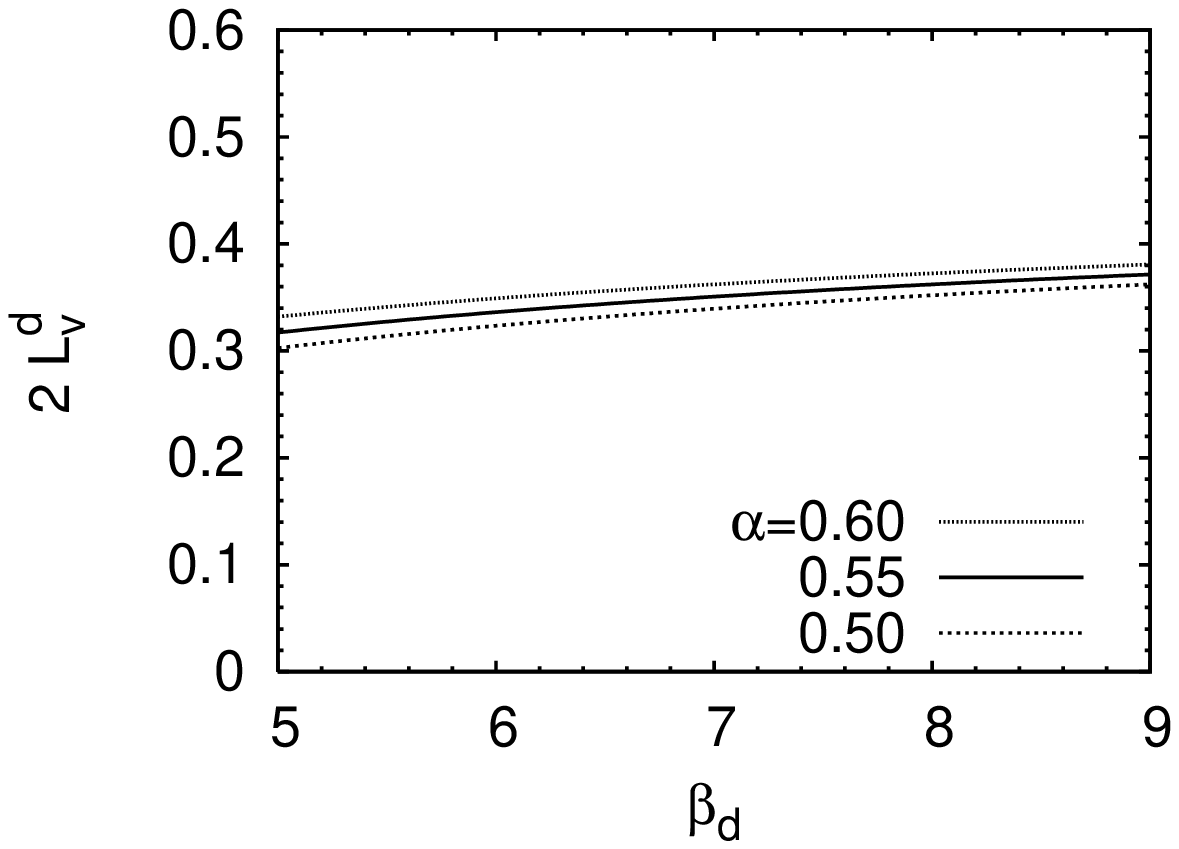}
\vspace{-3em}
\end{center}
\caption{\label{fig:L} The orbital angular momentum carried by valence
  quarks at scale $\mu=2 \gev$, obtained in the range of parameters
  for which we obtain a good fit to the proton and neutron Pauli form
  factors. The corresponding values for the total angular momentum
  $J_v^q = L_v^q + \half \int dx\, \Delta q_v(x) $ are $J_u = L_u +
  \half\, 0.93$ and $J_v^d = L_v^d - \half\, 0.34$, so that $J_v^d$
  comes out close to zero in our estimate.}
\end{figure*}

Sum rules analogous to (\ref{p-sum}) and (\ref{n-sum}) relate the
Pauli form factors of proton and neutron with the valence combinations
$E_v^q(x,t) = E^q(x,0,t) + E^q(-x,0,t)$ of proton helicity flip
distributions.  These distributions cannot be measured at $t=0$ so
that in contrast to $H_v^q(x,t)$ their forward limit is unknown.  We
have made an ansatz
\begin{equation}
  \label{E-ansatz}
E_v^q(x,t) = e_v^q(x) \exp[\, t g_q(x) \,]
\end{equation}
with $g_q(x)$ of the same functional form as $f_q(x)$ in
(\ref{profile-ansatz}).  For the forward limit of $E_v^q$ we assumed a
form
\begin{equation}
  \label{eq-ansatz}
e_v^q(x) = \mathcal{N}_q\, x^{-\alpha} (1-x)^{\beta_q} \,,
\end{equation}
which is known to work quite well for the ordinary valence quark
distributions $q_v(x)$.  The normalization factors $\mathcal{N}_q$ are
fixed by the requirement $\int dx\, e_v^q(x) = \kappa_q$ with
$\kappa_u\approx 1.67$ and $\kappa_d\approx -2.03$ obtained from the
magnetic moments of proton and neutron.  The overall normalization of
$E_v^q$ is hence quite large, which according to our discussion
implies significant spin-orbit effects of $u$ and $d$ quarks in the
proton.

Using the ansatz just described we obtain a good fit to the data for
the Pauli form factors of proton and neutron, with $\alpha=0.55$ and
$\alpha'= 0.9\gev^{-2}$ in agreement with expectations from Regge
phenomenology.  The quality of the fit is similar to the one we
achieved for the Dirac form factors.  We find a very large range of
allowed fit parameters, which is hardly surprising since we have to
determine both functions $e_v^q(x)$ and $g_q(x)$ in (\ref{E-ansatz}).
An important reduction of the allowed parameter space is due to the
positivity bound (\ref{e-bound}) and its analog in impact parameter
space at large $x$ (where it is reasonable to neglect the contribution
from antiquarks, which is subtracted in the valence distributions and
invalidates positivity conditions if it is large).  We find in
particular that $\beta_u \ge 3.5$ and $\beta_d \ge 5$ is required with
our ansatz, which quantifies our above statement that $E_v^q$ must
rather strongly decrease with~$x$.

We can now evaluate the orbital angular momentum carried by valence
quarks in the proton, which according to Ji's sum rule
\cite{Ji:1996ek} is
\begin{equation}
L_v^q =
\half \textstyle\int_0^1 dx\, 
  \Big[ x e_v^q(x) + x q_v(x) - \Delta q_v(x) \Big] .
\end{equation}
With our simple ansatz (\ref{eq-ansatz}) one readily obtains
\begin{equation}
{\textstyle\int_0^1} dx\, x e_v^q(x) = \kappa_q\,
  (1-\alpha) /(2-\alpha+\beta_q) .
\end{equation}
Remarkably, the results for $L_v^u$ and $L_v^d$ show only little
variation in the range of parameters $\alpha$, $\beta_u$ and $\beta_d$
for which we achieve a good fit to the Pauli form factors, as shown in
Fig.~\ref{fig:L}.  For the isovector combination we obtain a rather
well determined value $L_v^u - L_v^d = - \half\, ( 0.77 ~\mbox{to}~
0.92)$.  The isoscalar combination $L_v^u + L_v^d = - \half\, (0.11
~\mbox{to}~ 0.22)$ has a large uncertainty and is rather small, due to
partial cancellation between the two quark flavors.  Lattice
calculations by the QCDSF Collaboration obtain $L^u - L^d = - \half\,
(0.90\pm 0.12)$ and $L^u + L^d$ compatible with zero within errors
\cite{Schierholz:2005}.  We find the agreement with our estimates
encouraging, especially for the isovector combination, where
contributions from sea quarks should be small.

As in the case of $H_v^q$, we find that the different behavior of
$E_v^u$ and $E_v^d$ at large $x$ is reflected in a different $t$
dependence of their moments, as is characteristic for the Feynman
mechanism.  This is seen in the lower panels of
Fig.~\ref{fig:moments}, where we plot
\begin{equation}
  \label{e-moments}
e^q_{i}(t) = \textstyle \int_{0}^1 dx \, x^{i-1} \, E_v^q(x,t)
\end{equation}
for $i=1,2,3$ obtained with our best fit, where the large-$x$ powers
in (\ref{eq-ansatz}) are $\beta_u\approx 4$ and $\beta_d\approx 5.6$.
It will be very interesting to see whether such a behavior can be
established in form factor measurements or lattice calculations.

\begin{figure*}
\begin{center}
\includegraphics[width=0.49\textwidth]{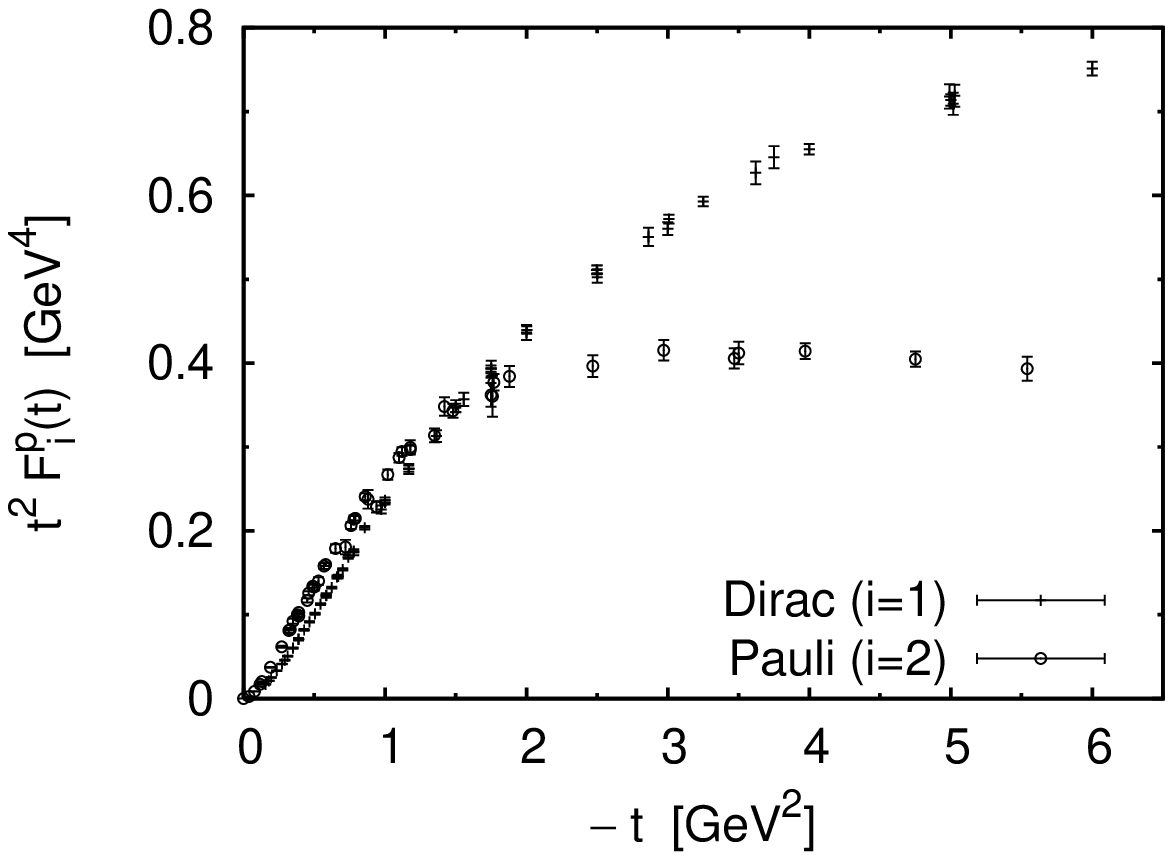}
\hfill
\includegraphics[width=0.49\textwidth]{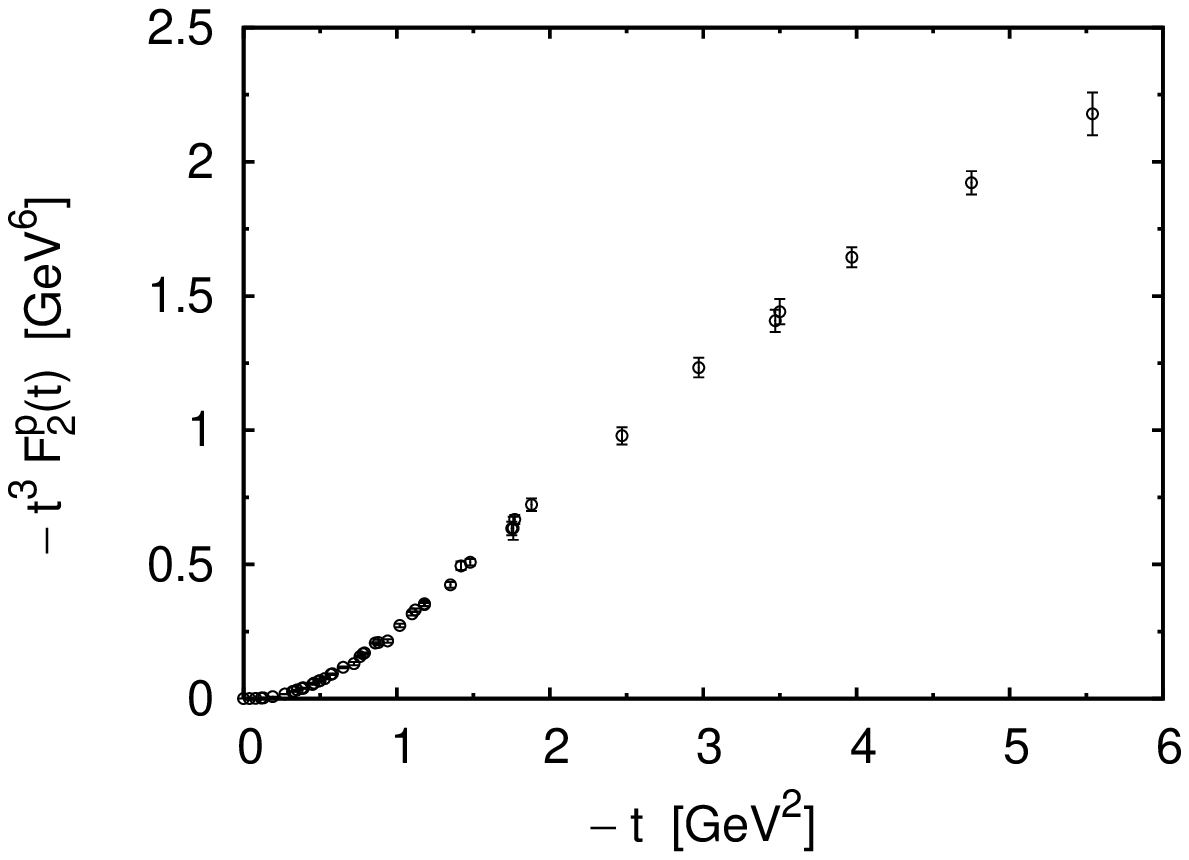}
\vspace{-3em}
\end{center}
\caption{\label{fig:F1F2} Data for the proton form factors, scaled by
  $t^2$ (left) or $\abst^3$ (right).  The plateau of $t^2 F_2^p(t)$
  for $\abst$ between $2$ and $6\gev^2$ illustrates that, in a limited
  kinematic region, observables may exhibit an approximate power law
  which is only transient and very different from their asymptotic
  behavior.}
\end{figure*}

Let us finally comment on the behavior of the ratio $F_2^p /F_1^p$ for
$\abst$ up to about $6\gev^2$, where there is data from the
polarization transfer method \cite{Gayou:2001qd}.  This data is rather
well described by a behavior $F_2^p /F_1^p \sim \abst^{-1}
\log^2(\abst/\Lambda^2)$ with $\Lambda\approx 300 \mev$, suggested by
the recent study \cite{Belitsky:2002kj} of $F_2^p$ in the
leading-twist hard-scattering mechanism \cite{Lepage:1980fj}.  For the
individual form factors, this study obtains an approximate behavior
\begin{eqnarray}
  \label{hsp}
F_1(t) &\sim & \alpha_s^{2 + 32/(9\beta)}\, \abst^{-2} ,
\nonumber \\
F_2(t) &\sim & \alpha_s^{2 + 8/(3\beta)}\,
               \log^2(\abst/\Lambda^2)\; \abst^{-3} ,
\end{eqnarray}
where the squared logarithm in $F_2$ arises from cutting off endpoint
singularities in the integrations over quark momentum fractions.  The
terms with $\beta=11-2n_f /3$ in the exponents are due to the
evolution of the proton distribution amplitude and numerically small,
with $32/(9\beta) \approx 0.4$ and $8/(3\beta) \approx 0.3$.  If one
takes $\alpha_s$ at scale $t$, then the logarithms in $F_2$
approximately cancel and $\abst^3 F_2(t)$ should be nearly $t$
independent.  If in contrast one assumes that the relevant scale in
$\alpha_s$ is so low that the running coupling is effectively frozen,
then $\abst^2 F_1(t)$ should be flat.  Figure~\ref{fig:F1F2} shows
that neither behavior is realized for $\abst$ below $6 \gev^2$, where
both $\abst^3 F_2^p(t)$ and $\abst^2 F_1^p(t)$ increase.  Thus, a
calculation of $F_1^p$ and $F_2^p$ at leading twist and leading order
in $\alpha_s$ not only underestimates the normalization of the form
factors (as remarked in \cite{Belitsky:2002kj}) but also fails to
describe their $t$ dependence in the region under discussion.  Whether
the fact that the ratio $F_2^p/F_1^p$ can be described with
(\ref{hsp}) is fortuitous or a sign of precious scaling behavior
remains a matter of debate.  An important source of power corrections
is the effect of transverse quark momentum in the hard-scattering
subprocess \cite{Bolz:1994hb}.  This effect decreases the form factors
more strongly at lower than at higher $\abst$, which may improve the
description of the $t$ dependence but makes the discrepancy for their
normalization worse.  The contribution of the Feynman mechanism to
$F_1^p$ and $F_2^p$ originates from different kinematic configurations
than the hard-scattering mechanism.  As a correction to the
leading-twist result it is hence additive rather than multiplicative
and will therefore not obviously cancel in the form factor ratio.

%%%%%%%%%%%%%%%%%%%%%%%%%%%%%%%%%%%%%%%%%%%%%%%%%%%%%%%%%%%%%

\section*{Acknowledgments}

It is a pleasure to thank D. Leinweber, L. von Smekal and T. Williams
for organizing an excellent workshop.  I am grateful to D. Br\"ommel
for a careful reading of the manuscript.  This work is supported by
the Helmholtz Association, contract number VH-NG-004.

%%%%%%%%%%%%%%%%%%%%%%%%%%%%%%%%%%%%%%%%%%%%%%%%%%%%%%%%%%%%%

\end{document}